\documentstyle[12pt]{article}
\title{ Space-like Penguin Effects in $B_{c}$ Decays 
        \thanks{Supported in part by National Natural Science 
	        Foundation of China } }
\author{ Dong-Sheng Du$\rm ^{a,b}$, 
         Zheng-Tao Wei$ \rm ^{b}$
         \thanks{E-mail address: duds@bepc3.ihep.ac.cn;$~~$
         $~~$weizt@hptc5.ihep.ac.cn} \\
         $\rm ^a$ \it{CCAST(World Laboratory),
         ~~P.O.Box 8730, Beijing 100080, China}\\
         $\rm ^b$ \it{Institute of High Energy Physics, Chinese
         Academy of Sciences,}\\
         \it{P.O.Box 918(4), Beijing, 100039, P. R. China}
         \thanks{mailing address.}}

\date{}
\begin{document}
\maketitle
\vspace*{0.3cm}


\section*{Abstract}

The space-like penguin contributions to branching ratios and CP asymmetries
in charmless decays of $B_{c}$ to two pseudoscalar mesons are studied using 
the next-to-leading order low energy effective Hamiltonian and factorization 
approximation. Both the gluonic penguin and the electroweak penguin 
diagrams are considered. In addition the annihilation diagram contributions 
are also taken in account. We find that the space-like penguin effects 
are significant.

\newpage
\section*{1. Introduction}

The weak decays of $B$ mesons offer a direct way to determine the 
Cabibbo-Kabayashi-Maskawa (CKM) matrix elements and to explore the 
origin of CP violation. Penguin diagrams can play an important role 
in charmless B decays. In most cases, attentions were paid to time-like 
penguin diagram in the literature, and the space-like penguin 
diagram is considered negligible because of the form factor suppression. 
In Ref.[1], space-like penguin diagram contribution to the 
branching ratios and CP violating asymmetries in $B_u^-$, $B^0$, 
$B_s^0$ decays are considered. The result shows that the space-like penguin 
amplitude can be enhanced 
by the hadronic matrix element involving (V-A)(V+A) or (S-P)(S+P) currents, 
and the space-like penguin effects are large in most charmless B decays.

$B_{c}$ meson is considered as the next and the last member of B
mesons. Its physics has got intensive attention recently [2][3]. $B_{c}$ decay
has its own characteristics. The obvious one is that $B_{c}$ 
carries $c$ and $b$ quarks, which are both heavy. So, $B_{c}$ decays can
be interesting candidates for testing the spectator ansatz. We assume that 
spectator approximation can be applied in $B_{c}$ decays. In
our paper, we will consider only $b$ quark decays and take $c$ quark
as a spectator. From Ref.[3], the future accelerator, Large Hardron Collider
(LHC), will produce $2.1\times 10^{8} B_{c}$ per year,  and can be a 
good place to study $B_{c}$ decays.

In this paper, we study space-like penguin diagram effects in $B_{c}$
decays to two pseudoscalars, and we concentrate on the charmless
$B_{c}$ decays, because penguin diagram plays an important role in
these decays. We use the next-to-leading order low energy effective
Hamiltonian and factorization approxamation to calculate the branching
ratios and CP violating asymmetries. In $B_{c}$ charmless decays, 
the annihilation diagram has the same order amplitude as the tree 
diagram. So the annihilation diagrams should be also taken into account.  
The result shows that the space-like penguin diagram 
contributions in $B_{c}$ charmless decays are large and can not be neglected.

\section*{2. Effective Hamiltonian and Factorization Approximation}

We assume spactator approximation in $B_{c}$ decays: the $c$ quark is 
a spectator and the $b$ quark decays to other light quarks. According to 
reference [4], the next-to-leading order low energy effective
Hamiltonian describing $|\Delta B|=1$ transitions
is given at the renormalization scale $\mu=O(m_b)$ as
$$
{\cal H}_{eff}(|\Delta B|=1) = \frac{G_F}{\sqrt{2}}
\left[\sum_{q=u,c}v_q \left\{ Q_1^qC_1(\mu)+Q_2^qC_2(\mu)
                       +\sum_{k=3}^{10} Q_k C_k(\mu)\right\}\right]+H.C.
\eqno(1)
$$

The CKM factors $v_q$ are defined as
$$
v_q=\left\{
          \begin{array}{ll}
          V_{qd}^*V_{qb} & \mbox{for $b \rightarrow d$ transitions}\\
          V_{qs}^*V_{qb} & \mbox{for $b \rightarrow s$ transitions.}
          \end{array}
     \right.
\eqno(2)
$$

The ten operators $Q_1^u$, $Q_2^u$, $Q_3,\ldots,Q_{10}$ are given as the
following forms:
$$
\begin{array}{ll}
Q_1^u=  (\bar{q}_{\alpha}u_{\beta})_{V-A}(\bar{u}_{\beta}b_{\alpha})_{V-A} &
Q_2^u=  (\bar{q}u)_{V-A}(\bar{u}b)_{V-A}\\
Q_{3(5)}=  (\bar{q}b)_{V-A}\displaystyle\sum_{q'}(\bar{q}'q')_{V-A(V+A)} &
Q_{4(6)}=  (\bar{q}_{\alpha}b_{\beta})_{V-A}
      \displaystyle\sum_{q'}(\bar{q}'_{\beta}q'_{\alpha})_{V-A(V+A)} \\
Q_{7(9)}=  \frac{3}{2}(\bar{q}b)_{V-A}
      \displaystyle\sum_{q'}e_{q'}(\bar{q}'q')_{V+A(V-A)} &
Q_{8(10)}=  \frac{3}{2}(\bar{q}_{\alpha}b_{\beta})_{V-A}
      \displaystyle\sum_{q'}e_{q'}(\bar{q}'_{\beta}q'_{\alpha})_{V+A(V-A)} \\
\end{array}
\eqno(3)
$$
where $Q_1^u$ and $Q_2^u$ are the current-current operators, and
the current-current operators $Q_1^c$ and $Q_2^c$ can be obtained from
$Q_1^u$ and $Q_2^u$ through the substitution of $u\rightarrow c$.
$Q_3,\ldots,Q_6$ are the QCD penguin operators, whereas $Q_7,\ldots,Q_{10}$
are the electroweak penguin operators. The quark $q=d~or~s$  for
$b \rightarrow
d~or~s$ transitions, respectively;  the indices $\alpha,~\beta$
are $SU(3)_c$ color indices; $(V \pm A)$ refer to $\gamma_{\mu}(1\pm
\gamma_5)$.  

It is useful to  use the renormalization scheme independent Wilson 
coefficient functions
[5]:
$$
{\bf \bar{C}}(\mu)=\left[\hat{1}+\frac{\alpha_s(\mu)}{4\pi}\hat{r}_s^T
                    +\frac{\alpha(\mu)}{4\pi}\hat{r}_e^T\right]
                    \cdot {\bf C}(\mu),
\eqno(4)
$$
where ${\bf C}(\mu)$, ${\bf \bar{C}}(\mu)$ 
are all column vectors.  The matrix elements are: 
$$
<{\bf Q}^T(\mu)\cdot {\bf C}(\mu)>
\equiv<{\bf Q}^T>_0\cdot{\bf C'}(\mu)
\eqno(5)
$$
where $<{\bf Q}>_0$ denote
the tree level matrix elements of these operators,
and  ${\bf C'}(\mu)$ are defined as
$$
\begin{array}{llll}
C'_1=~\overline{C}_1, &C'_2~=~\overline{C}_2, &
C'_3=~\overline{C}_3-P_s/3, &C'_4~=~\overline{C}_4+P_s,\\
C'_5=~\overline{C}_5-P_s/3, & C'_6~=~\overline{C}_6+P_s,&
C'_7=~\overline{C}_7+P_e,   &C'_8~=~\overline{C}_8, \\
C'_9=~\overline{C}_9+P_e,   &C'_{10}~=~\overline{C}_{10},
\end{array} \eqno(6)
$$
where $P_{s,e}$ are given by
$$
\begin{array}{rl}
P_s&=~\frac{\alpha_s}{8\pi}\overline{C}_2(\mu)\left[\frac{10}{9}-
       G(m_q,q,\mu)\right],\\
P_e&=~\frac{\alpha_{em}}{9\pi}\left(3\overline{C}_1+\overline{C}_2(\mu)
      \right)\left[\frac{10}{9}-G(m_q,q,\mu)\right],\\
G(m,q,\mu)&=~-4\int_0^1 dx~x(1-x)ln\displaystyle\left[\frac{m^2-x(1-x)q^2}
{\mu^2}\right],
\end{array} \eqno(7)
$$
here $q=u,~c$.
The numerical values of the renormalization scheme
independent Wilson Coefficients $\overline{C}_i(\mu)$ at $\mu=O(m_b)$
are [6]
$$
\begin{array}{llll}
\bar{ c}_1=-0.313, & \bar {c}_2=1.150, &\bar{ c}_3=0.017, &
\bar{ c}_4=-0.037, \\
\bar {c}_5=0.010,  & \bar{ c}_6=-0.046, &
\bar{ c}_7=-0.001\cdot \alpha_{em}, \\
\bar {c}_8=0.049\cdot\alpha_{em}, &
\bar{ c}_9=-1.321\cdot\alpha_{em}, &
\bar{ c}_{10}=~0.267\cdot\alpha_{em}.
\end{array}\eqno(8)
$$

In Equation (7), $q^2$ denotes the momentum transfer squared of the virtual 
gluons, photons, and $Z^0$ appearing in the QCD and electroweak penguin 
didagrams respectively. So, the Wilson coefficients $C_{i}'$ depend on $q^2$. 
We adopt a simple kinematic picture [1] for two body 
decays $B\rightarrow PP'$ as illustrated as in Fig 1. 

The average value of $q^2$ can be given by
$$  
   <q^2>=m_{b}^{2}+m_{q}^{2}-2m_{b}E_{q}
\eqno(9)
$$
where $E_{q}$ is determined from 
$$
  E_{q}+\sqrt{E_{q}^{2}-m_{q}^{2}+m_{q'}^{2}}+
        \sqrt{4(E_{q}^{2}-m_{q}^{2})+m_{q'}^{2}}=m_{b}
\eqno(10)
$$
for the time-like penguin diagram; and 
$$
  E_{q}+\sqrt{E_{q}^{2}-m_{q}^{2}+m_{q'}^{2}}+=m_{b}+m_{q'}
\eqno(11)
$$
for the space-like penguin diagram.

\vspace{1cm}

In exclusive nonleptonic decays, the current-current operator 
matrix element can be calculated by factorization approximation 
and BSW method [7].

For the tree diagram of Fig.2a which correponds to 
$b\rightarrow q\bar{q'}q'$, the matrix element for four-quark 
operator is defined as:
$$
\begin{array}{ll}
M^{PP'}_{q_{1}q_{2}q} &\equiv<PP'|(\bar{q_1}q_2)_{V-A}
               (\bar{q}b)_{V-A}|B_{c}^{-}>\\
               &=<P|(\bar{q_1}q_2)_{V-A}|0><P'|(\bar{q}b)_{V-A}|B_{c}^{-}>\\
	       &=-if_{P}^{\bar{q_1}q2}f_{+}^{B_{c}P'}(M_{P}^2)(M_{B_c}^2-
	       M_{P'}^2-\frac{M_{B_c}-M_{P'}}{M_{B_c}+M_{P'}}M_{X}^2) 
\end{array}\eqno(12)
$$
where $f_{+}^{B_{c}P'}(m_{P}^2)=\frac{f_{+}^{B_{c}P'}(0)}
{1-M_{P}^2/(M_{B_c}^{pole})^2}$. $ f_{+}^{B_{c}P'}(0) $ can be calculated
in BSW model, and $ M_{B_c}^{pole}=6.30GeV $. For the time-like penguin 
diagram, this factorization method is 
applied to calculate the time-like penguin operator matrix element.

For the annihilation diagram of Fig.2b corresponds to 
$b\bar{c} \rightarrow q\bar{c}$, the matrix element is [8]: 
$$
\begin{array}{ll}
S^{PP'}_{qcc} &\equiv<PP'|(\bar{q}c)_{V-A}(\bar{c}b)_{V-A}|B_{c}^{-}>\\
               &=<PP'|(\bar{q}c)_{V-A}|0><0|(\bar{c'}b)_{V-A}|B_{c}^{-}>\\
	       &=if_{B_c}f_{+}^{a}(M_{B_c}^2)(M_{P}^2-
	       M_{P'}^2-\frac{M_{P}-M_{P'}}{M_{P}+M_{P'}}M_{B_c}^2) 
\end{array}\eqno(13)
$$
The matrix elements are computed at momentum transfer $q^2=M_{B_c}^2$.
We take the asymptotic form factor $f_+^a(M_{B_c}^2)=
i16\pi \alpha_s f_{B_c}^2/M_{B_c}^2$ [9].
One point should be noted: for the annihilation diagram,
$C_{1}'=\bar{C_2}$, $C_{2}'=\bar{C_{1}}$ .

In $B_c$ decays, the annihilation diagram is enhanced by the
CKM factor $v_c$. For the $b\rightarrow d$ process,
$|\frac{v_c}{v_u}|\approx 3$; for $b\rightarrow s$ process, 
$|\frac{v_c}{v_u}|\approx 57$. So, the annihilation diagram should be 
taken into account in $B_c$ decays.
 
For the space-like penguin diagram, just like the annihilation diagram, 
its factorization method is the same as that of the annihilation 
diagram.

\section*{3. Numerical Calculation}

The decay width for a $B_c$ meson at rest decaying into two 
pseudoscalars is
$$
 \Gamma(B_c \to PP^{'})=
          \frac{1}{8\pi}|<PP^{'}|H_{eff}|B_c>|^2\frac{|\stackrel{
	  \rightarrow}{p}|}{M_{B_c}^2}
\eqno(14)
$$
where
$$
  |\stackrel{\rightarrow}{p}|=\frac{[(M_{B_c}^2-(M_P+M_P{'})^2)
  (M_{B_c}^2-(M_P-M_P{'})^2)]^{\frac{1}{2}}}{2M_{B_c}}
\eqno(15)
$$
is the momentum of the pseudoscalar meson $P$ or $P'$.
The corresponding branching ratios are given by
$$
Br(B_c\to PP^{'})=\frac{\Gamma(B_c \rightarrow PP^{'})}{\Gamma_{tot}^{B_c}}.
\eqno(16)
$$
In our numerical calculation, we take[10]
$\Gamma_{tot}^{B_c}=1.32\times10^{-12}$GeV,

The $B_c$ meson decay amplitude can be generally expressed as
$$
<PP'|H_{eff}|B_c^->=\frac{G_F}{\sqrt{2}}\sum_{q=u,c}v_q F_q.
\eqno(17)
$$
where $q=u,c$, and $F_q$ including the tree and and annihilation 
and penguin amplitude.  

The CP-violating asymmetry  can be given by
$$\begin{array}{rl}
\displaystyle{\cal A}_{cp}
&\equiv~\displaystyle\frac{\Gamma(B_c^- \rightarrow PP^{'})-
       \Gamma(B_c^+ \rightarrow \bar{P}\bar{P{'}})}
      {\Gamma(B_c^- \rightarrow PP^{'})+
       \Gamma(B_c^+ \rightarrow \bar{P}\bar{P{'}})}\\[4mm]
&=~\displaystyle\frac{2Im(v_uv_c^*)Im(F_c/F_u)}{|v_u|^2+
    |v_c|^2|F_c/F_u|^2+2Re(v_uv_c^*)Re(F_c/F_u)}.
\end{array} \eqno(18)
$$

We take the decay $B_c^-\rightarrow \eta D^- $ as an example to
illustrate the calculation of branching ratio $Br$ and CP asymmetry 
${\cal A}_{cp}$ including space-like penguin diagram.
$$\begin{array}{ll}
<\eta D^-|H_{eff}|B_c^->=
&\frac{G_F}{\sqrt{2}}\sum\limits_{q=u,c} v_q [(a_2\delta_{uq}
+a_3+a_4-a_6+a_8-a_{9}/2-a_{10}+\\
&\frac{2M_{\eta}^2}{(m_d+m_d)(m_b-m_d)}(a_5-a_7/2))
M^{\eta D^{-}}_{ddd}+\\
&(a_2\delta_{cq}+a_3+\frac{2M_{B_c}^2}{(m_d-m_c)(m_b+m_c)}(a_5+a_7)+a_9)
S^{\eta D^-}_{dcc} ].
\end{array}\eqno(19)
$$

where  $a_k$ is defined as
$$\begin{array}{ll}
a_{2i-1}&\equiv~\displaystyle\frac{C'_{2i-1}}{3}+C'_{2i},\\[4mm]
a_{2i}&\equiv~ C'_{2i-1}+\displaystyle\frac{C'_{2i}}{3},~(i=1,2,3,4,5)
\end{array}
$$
and 
$$\begin{array}{ll}
M_{ddd}^{\eta D^-}=-i f^{\bar{d}d}_{\eta}f_+^{B_c^-D^-}(M_{\eta}^2)\left[
      (M_{B_c}^2-M_{D^-}^2)-\frac{M_{B_c}-M_{D^-}}
      {M_{B_c}+M_{D^-}}M_{\eta^2}\right] \\[4mm]
S_{dcc}^{\eta D^-}=\frac{i}{\sqrt{3}} f_{B_c}f_+^a(M_{B_c}^2)\left[
      (M_{\eta}^2-M_{D^-}^2)-\frac{M_{\eta}-M_{D^-}}
      {M_{\eta}+M_{D^-}}M_{B_c}^2\right] 
\end{array} \eqno(20)
$$
where $\frac{1}{\sqrt{3}}$ arises from $\eta=\frac{\bar{u}u+
\bar{d}d-\bar{s}s}{\sqrt{3}}$. The $a_{2}\delta_{cq}$ term 
in Eq.(19) is the annihilation diagram contribution.

The numerical results of the space-like penguin contributions to the
branching ratios and CP-violating asymmetries are given in Table 1 and 2. 
We calculate the branching ratios and CP-violating asymmetries 
with the tree and annihilation and time-like penguin
contributions for comparison.  All the parameters such as
meson decay constants, form factors and quark masses needed in our
calculation are taken as
$f_{\pi^{\pm}}=0.13GeV$,
$f_K=0.160GeV$[11], $f_{\pi^0}^{{\bar u}u}=-f_{\pi^0}^{\bar {d}d}=f_{\pi^\pm}
/\sqrt{2}$.
$f_{\eta}^{\bar {u}u}=f_{\eta}^{\bar {d}d}=-f_{\eta}^{\bar {s}s}=0.077GeV$,
$f_{\eta^{'}}^{\bar {u}u}=f_{\eta^{'}}^{\bar {d}d}=f_{\eta{'}}^{\bar
{s}s}/2=0.054GeV$[12],
$f_{B_c}=0.5GeV$[3],
$f_+^{B_c^-D^-}(0)=0.595$, $f_+^{B_c^-D_s^-}(0)=0.605$,
$m_u=0.005GeV$,
$m_d=0.01GeV$,
$m_s=0.2GeV$,
$m_c=1.5GeV$,
$m_b=4.5GeV$,
$M_{B_c}=6.27GeV$.
and the Wolfenstein parametrized CKM parameters are [13]:
$\lambda=0.22$,
$A=0.8$,
$\eta=0.34$,
$\rho=-0.12$.

\section*{4. Conclusion and discussion }

{}From Table 1 and 2 we can see the following features:

(i) For most of the charmless decays, space-like penguin contributions
to branching ratios are large. The corrections to the branching ratio
and CP violating asymmetries are more than $100\%$.

(ii) For space-like penguin in $B_c$ decays, the contributions of the 
electro-weak penguins are negligible.

(iii) The annihilation diagram contribution can not be negligible 
in $B_c$ decays.  

The reason for the large space-like penguin effects can be explained 
as follows:

(i) When calculating the matrix elememt of $(V-A)(V+A)$ current 
$<PP'|(\bar{q}b)_{V-A}(\bar{c}c)_{V+A}|B_c>$, there will 
appear a factor $\frac{2m_{B_c}^2}{(m_q-m_c)(m_b+m_c)}$, this 
factor will enhance the  space-like penguin effects.

(ii) The form factor $f_+^a(m_{B_c}^2)$ is not a suppression factor
as usually considered. In $B_c$ decays, $f_+^a(m_{B_c}^2)=0.077$, so
combine with $f_{B_c}=0.5 $, the annihilation or space-like penguin
matrix element $S^{PP'}$
is nearly as that of the tree and time-like penguin matrix element 
$M^{PP'}$.   

(iii) The quark mass is an important and sensitive parameter. In our 
calculation, we have used the current mass. The values 
of quark mass will have direct effect on the penguin amplitude. 
Other effects, such as nonfactorization effect, final state 
interation can provide many uncertainties.

\section*{Acknoledgement}

This work is supported in part by National Natural
Science Foundation of China and the Grant of State 
Commission of Science and Technology of China.

\newpage
\section*{Figure Captions}

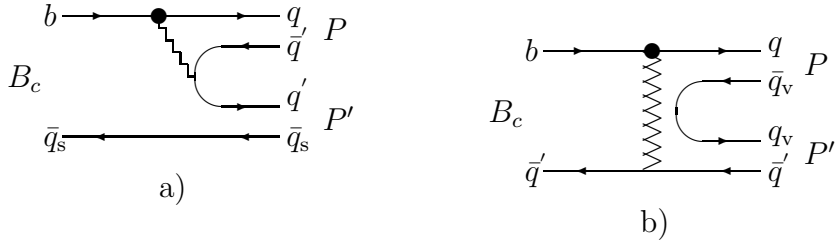
Fig. 1. Penguin diagrams for a $B_c$  meson decaying into two light
pseudoscalar mesons $P$ and $P{'}$ a) the time-like penguin diagram; 
b) the space-like penguin diagram. 
The subscripts ``v" denote ``vacuum".
The dark dot stands for the contraction of the W-loop.\\

\noindent
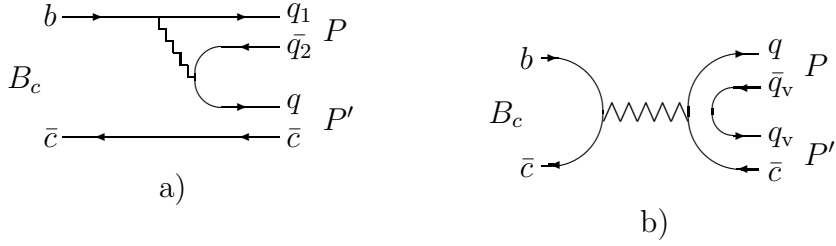
Fig. 2. The Tree diagram and the Annihilation diagram in $B_c$ decays. 
a) the Tree diagram. b) the Annihilation diagram. 

\newpage

\begin{figure}
\unitlength=0.32mm

\begin{picture}(0,275)
\put(110,255){\circle*{7}}
\put(70,255){\line(1,0){90}}
\put(70,205){\line(1,0){90}}
\put(62,251){$b$}
\put(62,201){$\bar{q}_{\rm s}$}
\put(47,225.5){$B_c$}
\put(163,253){$q$}
\put(163,201){$\bar{q}_{\rm s}$}
\put(163,239){$\bar{q}^{'}$}
\put(163,218){$q^{'}$}
\put(178,208){$P{'}$}
\put(178,245){$P$}
\put(160,230){\oval(70,25)[l]}
\put(85,255){\vector(1,0){2}}
\put(85,205){\vector(-1,0){2}}
\put(145,255){\vector(1,0){2}}
\put(145,205){\vector(-1,0){2}}
\put(145,242.5){\vector(-1,0){2}}
\put(145,217.3){\vector(1,0){2}}
\multiput(110,255)(3,-5){5}{\line(0,-1){5}}
\multiput(107,255)(3,-5){6}{\line(1,0){3}}
\put(110,180){a)}
\end{picture}

\begin{picture}(0,-200)
\put(315,270){\circle*{7}}
\put(270,270){\line(1,0){90}}
\put(270,220){\line(1,0){90}}
\put(262,266){$b$}
\put(262,216){$\bar{q}^{'}$}
\put(247,240.5){$B_c$}
\put(363,270){$q$}
\put(363,216){$\bar{q}^{'}$}
\put(363,254){$\bar{q}_{\rm v}$}
\put(363,233){$q_{\rm v}$}
\put(378,223){$P{'}$}
\put(378,260){$P$}
\put(360,245){\oval(70,25)[l]}
\put(285,270){\vector(1,0){2}}
\put(285,220){\vector(-1,0){2}}
\put(345,270){\vector(1,0){2}}
\put(345,220){\vector(-1,0){2}}
\put(345,257.5){\vector(-1,0){2}}
\put(345,232.5){\vector(1,0){2}}
\multiput(310,263)(0,-6){8}{$>$}
\put(310,195){b)}
\put(200, 165){Fig. 1.}
\end{picture}

\begin{picture}(350,0)
\put(70,105){\line(1,0){90}}
\put(70,55){\line(1,0){90}}
\put(62,101){$b$}
\put(62,51){$\bar{c}$}
\put(47,75.5){$B_c$}
\put(163,90){$\bar{q_2}$}
\put(163,51){$\bar{c}$}
\put(163,105){$q_1$}
\put(163,67){$q$}
\put(178,58){$P{'}$}
\put(178,95){$P$}
\put(160,80){\oval(70,25)[l]}
\put(85,105){\vector(1,0){2}}
\put(85,55){\vector(-1,0){2}}
\put(145,105){\vector(1,0){2}}
\put(145,55){\vector(-1,0){2}}
\put(145,92.5){\vector(-1,0){2}}
\put(145,67.3){\vector(1,0){2}}
\multiput(110,105)(3,-5){5}{\line(0,-1){5}}
\multiput(107,105)(3,-5){6}{\line(1,0){3}}
\put(110,30){a)}
\end{picture}

\begin{picture}(350,0)
\put(260,113){$b$}
\put(260,68){$\bar{c}$}
\put(247,90.5){$B_c$}
\put(363,119.5){$q$}
\put(363,66){$\bar{c}$}
\put(363,103.7){$\bar{q}_{\rm v}$}
\put(363,83){$q_{\rm v}$}
\put(378,73){$P{'}$}
\put(378,110){$P$}

\put(269,95){\oval(51,45)[r]}
\put(359,95){\oval(58,48)[l]}
\put(360,95){\oval(40,20)[l]}
\put(274,117.3){\vector(1,0){2}}
\put(274,73){\vector(-1,0){2}}
\put(352,119.1){\vector(1,0){2}}
\put(352,71.3){\vector(-1,0){2}}
\put(352,104.9){\vector(-1,0){2}}
\put(352,84.8){\vector(1,0){2}}
\multiput(294,91)(7,0){5}{$\wedge$}
\put(310,45){b)}
\put(200, 15){Fig. 2.}
\end{picture}
\end{figure}

\newpage

\section*{Table Captions}

Table 1. The Branching Ratios of $B_c$ decaying to two pseudoscalar. \\ 
Table 2. The CP Asymmetries of $B_c$ decaying to two pseudoscalar.  
where the ``Tree" means the  tree diagram contribution, 
``Anni'' means the annihilation diagram contribution, 
``T-like" denotes the time-like penguin contributions, the ``S-like"
denotes the space-like penguin contributions, ``QCD" means QCD penguin 
contributions, and ``EW" means 
electro-weak penguin contributions.

\newpage

\begin{center}
Table 1\\
\vspace{0.5cm}
\begin{scriptsize}
\begin{tabular}{|c|c|c|c|c|c|c|} \hline

 Decay Mode & \multicolumn{6}{|c|} { Br } \\\cline{2-7}
  & Only Tree & Tree+Anni & \multicolumn{2}{|c|} {Tree+Anni+T-like} &
    \multicolumn{2}{|c|} {Tree+Anni+T-like+S-like}\\ \cline{4-7}  
  &  &  & QCD & QCD+EW & QCD  & QCD+EW   \\\hline
  
$B_c^- \rightarrow \pi^- \bar{D^0}$ &
$1.12\times10^{-5}$ & $9.60\times10^{-6}$ & $8.31\times10^{-6}$ &  
$8.29\times10^{-6}$ & $2.27\times10^{-6}$ & $2.32\times10^{-6}$ \\\hline

$B_c^- \rightarrow K^- \bar{D^0}$ &
$8.63\times10^{-7}$ & $2.81\times10^{-6}$ & $1.94\times10^{-5}$ &  
$1.99\times10^{-5}$ & $4.82\times10^{-5}$ & $4.76\times10^{-5}$ \\\hline

$B_c^- \rightarrow \pi^0 D^-$ &
$2.54\times10^{-8}$ & $1.20\times10^{-7}$ & $5.21\times10^{-7}$ &  
$3.73\times10^{-7}$ & $2.01\times10^{-5}$ & $1.91\times10^{-5}$ \\\hline

$B_c^- \rightarrow \eta D^-$ &
$1.77\times10^{-8}$ & $3.00\times10^{-9}$ & $8.05\times10^{-6}$ & 
$7.73\times10^{-6}$ & $7.09\times10^{-6}$ & $7.10\times10^{-6}$ \\\hline

$B_c^- \rightarrow \eta' D^-$ &
$1.76\times10^{-8}$ & $9.88\times10^{-9}$ & $6.45\times10^{-5}$ & 
$6.43\times10^{-5}$ & $6.34\times10^{-5}$ & $6.34\times10^{-5}$ \\\hline

$B_c^- \rightarrow \eta D_{s}^-$ &
$9.03\times10^{-10}$ & $1.53\times10^{-7}$ & $8.62\times10^{-6}$ & 
$2.71\times10^{-6}$  & $9.02\times10^{-6}$ & $8.87\times10^{-6}$ \\\hline

$B_c^- \rightarrow \eta' D_{s}^-$ &
$4.40\times10^{-10}$ & $1.26\times10^{-7}$ & $6.09\times10^{-6}$ & 
$9.18\times10^{-6}$  & $1.18\times10^{-5}$ & $1.17\times10^{-5}$ \\\hline

$B_c^- \rightarrow K^0 D_{s}^-$ &
   0                 & $3.41\times10^{-8}$ & $1.29\times10^{-6}$ & 
$1.27\times10^{-6}$  & $1.99\times10^{-6}$ & $1.96\times10^{-5}$ \\\hline

$B_c^- \rightarrow \bar{K^0} D^-$ &
  0                 & $5.87\times10^{-7}$ & $1.60\times10^{-5}$ & 
$1.58\times10^{-5}$  & $3.60\times10^{-5}$ & $3.55\times10^{-5}$ \\\hline

\end{tabular}

\end{scriptsize}

\vspace{2cm}

Table 2\\

\vspace{0.5cm}
\begin{tabular}{|c|c|c|c|c|c|} \hline
 Decay Mode & \multicolumn{5}{|c|} { ${\cal{A}}_{cp}$ } \\\cline{2-6}
  & Tree+Anni & \multicolumn{2}{|c|} {Tree+Anni+T-like} &
    \multicolumn{2}{|c|} {Tree+Anni+T-like+S-like}\\ \cline{3-6}  
  &  & QCD & QCD+EW & QCD  & QCD+EW   \\\hline

$B_c^- \rightarrow \pi^- \bar{D^0}$ &
$-15.1\%$ & $-6.8\%$ & $-6.8\%$ & $-80.6\%$ & $-80.1\%$ \\\hline 

$B_c^- \rightarrow K^- \bar{D^0}$ &
$92.6\%$ & $-0.7\%$ & $-0.7\%$ & $19.7\%$ & $19.7\%$ \\\hline 

$B_c^- \rightarrow \pi^0 D^-$ &
$92.7\%$ & $27.9\%$ & $34.1\%$ & $15.1\%$ & $15.1\%$ \\\hline 

$B_c^- \rightarrow \eta D^-$ &
$-88.6\%$ & $14.8\%$ & $15.1\%$ & $9.2\%$ & $9.3\%$ \\\hline 

$B_c^- \rightarrow \eta' D^-$ &
$-47.7\%$ & $12.1\%$ & $12.1\%$ & $11.6\%$ & $11.6\%$ \\\hline 

$B_c^- \rightarrow \eta D_{s}^-$ &
$-13.4\%$ & $-0.7\%$ & $-1.7\%$ & $-1.4\%$ & $-1.4\%$ \\\hline 

$B_c^- \rightarrow \eta' D_{s}^-$ &
$11.8\%$ & $-1.8\%$ & $-1.4\%$ & $0.3\%$ & $0.3\%$ \\\hline 

$B_c^- \rightarrow K^0 D_{s}^-$ &
0        & $20.2\%$ & $20.4\%$ & $-0.6\%$ & $-0.6\%$ \\\hline 

$B_c^- \rightarrow \bar{K^0} D^-$ &
0        & $-1.3\%$ & $-1.3\%$ & $0.1\%$ & $0.1\%$ \\\hline

\end{tabular}

\end{center}

\end{document}